\documentclass[letterpaper, conference, twocolumn, 11pt]{IEEEtran}

\usepackage[utf8]{inputenc}
\usepackage{graphicx}
\usepackage{amssymb,amsthm} 
\usepackage[cmex10]{amsmath}
\usepackage{xcolor}
\usepackage{xifthen}
\usepackage{pifont}
\usepackage{hyperref}
\hypersetup{colorlinks=false,linkbordercolor=red,citebordercolor=blue,pdfborderstyle={/S/U/W 1}}
\usepackage{url}
\usepackage[english]{babel}
\usepackage[english = american]{csquotes}
\usepackage[final]{pdfpages}
\usepackage{enumitem}
\usepackage[ruled,vlined, linesnumbered]{algorithm2e}

\SetCommentSty{mycommfont}

\newtheorem{definition}{Definition}
\newtheorem{lemma}[definition]{Lemma}
\newtheorem{theorem}[definition]{Theorem}

\newtheorem{corollary}[definition]{Corollary}
\newtheorem{notation}[definition]{Notation}

\newcommand\mydef{\mathrel{\overset{\makebox[0pt]{\mbox{\normalfont\tiny\sffamily def}}}{=}}}
\newcommand{\qr}{\texttt{QR}}

\newcommand{\Fq}[1]{\ensuremath{\mathbb{F}_{#1}}}
\newcommand{\inter}[2]{
  \ifthenelse{\equal{#2}{\empty}}
    { \ensuremath{[#1]} }
    { \ensuremath{[#1, #2]} }
}
\newcommand{\LIN}[4]{\ensuremath{[#1,#2,#3]_{#4}}}
\newcommand{\pp}[1]{\ensuremath{pp_{#1}}}
\newcommand{\parity}[1]{\ensuremath{p(#1)}}
\newcommand{\parityd}[1]{\ensuremath{p_d(#1)}}
\newcommand{\cv}[2]{\ensuremath{cv(#1,#2)}}

\newcommand{\const}[1]{\ensuremath{\mathsf{cst}_{#1}}}
\newcommand{\key}[1]{\ensuremath{\mathsf{key}_{#1}}}
\newcommand{\ctr}[1]{\ensuremath{\mathsf{ctr}_{#1}}}
\newcommand{\nonce}[1]{\ensuremath{\mathsf{nonce}_{#1}}}
\newcommand{\azero}{\ensuremath{a^{0}}}
\newcommand{\bzero}{\ensuremath{b^{0}}}
\newcommand{\czero}{\ensuremath{c^{0}}}
\newcommand{\dzero}{\ensuremath{d^{0}}}
\newcommand{\bone}{\ensuremath{b^{1}}}
\newcommand{\btwo}{\ensuremath{b^{2}}}
\newcommand{\done}{\ensuremath{d^{1}}}
\newcommand{\dtwo}{\ensuremath{d^{2}}}

\renewcommand{\tilde}{\widetilde}
\renewcommand{\bar}{\overline}

\newcommand{\myalgo}[1]{\begin{algorithm}[h] {\footnotesize #1 } \end{algorithm}}

\begin{document}

\title{Parity-Based Concurrent Error Detection Schemes for the ChaCha Stream Cipher}

\IEEEoverridecommandlockouts

\author{\IEEEauthorblockN{Viola Rieger and Alexander Zeh}
\IEEEauthorblockA{Research and Development Center \\ Infineon Technologies AG, Munich, Germany\\
\texttt{ \{viola.rieger, alexander.zeh\}@infineon.com}}}

\maketitle

\begin{abstract}
We propose two parity-based concurrent error detection schemes for the Quarterround of the ChaCha stream cipher to protect from transient and permanent faults. 
They offer a trade-off between implementation overhead and error coverage. 
The second approach can detect any odd-weight error on the in-/output and intermediate signals of a Quarterround, while the first one requires less logic.
\end{abstract}

\section{Introduction}
The ChaCha stream cipher was introduced by Bernstein in 2008 \cite{bernstein_chacha_2008} as a successor of the Salsa cipher family \cite{bernstein_salsa20_2008}. Both algorithms are based on pseudorandom functions using ADD, ROTATE and XOR (ARX) operations. 
In order to provide confidentiality, authenticity and integrity for data, Authenticated Encryption with Associated Data (AEAD) takes on greater significance and realization in hardware for embedded systems seems fruitful.  
With the authenticator Poly1305, Bernstein proposed a software optimized Message Authentication Code (MAC, \cite{bernstein_poly1305-aes_2005}), that is well suited to be used in combination with the ChaCha algorithm. 
Concurrent Error Checking (CED, \cite{liden_latching_1994,goessel_new_2008}) was developed to detect faults within functional or logical building blocks of embedded circuits, such as ALUs, adders or individual gates. CED enables an efficient, testable and robust design. Parity-based CED was applied to substitution-permutation networks in~\cite{karri_parity-based_2003} and, e.g., applied to AES
by Bertoni~\cite{bertoni_error_2003} and Natale~\cite{natale_novel_2007}. To our knowledge, no CED scheme has been proposed for ChaCha so far.\\
This paper is structured as follows: we recall preliminaries on parity codes and the parity-bit prediction for basic operations on binary vectors in Section~\ref{sec_Preliminaries}.
Section~\ref{sec_ChaCha} contains basic operations of the ChaCha-Poly1305 AEAD scheme and intermediate signals in a Quarterround of ChaCha are defined. The transformation of a Quarterround into a code-disjoint circuit based on a single parity-bit prediction is described in Section~\ref{sec_Classic}. Thm.~\ref{theo_PP-ChaCha} gives the expression for the overall parity-bit and the error coverage is proven in Thm.~\ref{theo_ERR-PP-ChaCha}. Our second group-based parity prediction is described in Section~\ref{sec_GroupBasedPP} and Thm.~\ref{theo_OddBitErrors} proofs its error coverage. 


\section{Preliminaries} \label{sec_Preliminaries}
Let $\Fq{q}$ denote the finite field of order $q$. For two integers $a, b$ with $b > a$, we denote by $\inter{a}{b}$ the integer set $\{i \in \mathbb{Z} : a \leq i \leq b \}$ and let $\inter{b}{}$ be the shorthand notation for $\inter{1}{b}$. 
Similarly, $\LIN{n}{k}{d}{q}$ denotes the parameters of a $q$-ary \emph{linear} code of length $n$, dimension $k$, and minimum Hamming distance $d$. A generator matrix of a linear $\LIN{n}{k}{d}{q}$ code $\mathcal{C}$ over $\Fq{q}$ is a $k \times n$ matrix whose rows form a basis of $\mathcal{C}$.
The binary $\LIN{n}{n-1}{2}{2}$ parity code is defined as the code with generator matrix $G = (I \ 1_{n-1})$, where $I$ is the $(n-1) \times (n-1)$ identity matrix and $1_{n-1}$ is $(1 \ 1 \ \cdots \ 1)^T$.
It is well-known, that any error $e \in \Fq{2}^n$ of odd Hamming weight added to a codeword $c$ of an $\LIN{n}{n-1}{2}{2}$ parity code $\mathcal{C}$ results in a vector $(c+e) \not\in \mathcal{C}$ that is detectable. In the following we use the calculation of the one-bit redundancy of a parity code. 
The bitwise XOR of two binary vectors $a,b \in \Fq{2}^n$ is denoted by $a \oplus b$ and $a \boxplus b$ is the addition of $a$ and $b$ in $\Fq{2}^n$.\\
A parity-bit of $x = (x_0 \ x_1 \ \cdots \ x_{n-1}) \in \Fq{2}^n$ is defined as the Boolean function $
p:  \Fq{2}^n   \mapsto  \Fq{2}$, where $\parity{x} \mydef  \bigoplus_{i=0}^{n-1} x_i$. Let $a,b \in \Fq{2}^n$, then we have:
\begin{eqnarray} \label{eq_PPDirectSum}
\parity{a \oplus b} & = & \parity{a} \oplus \parity{b}.
\end{eqnarray}
The parity-bit of the sum of two binary vectors is: 
\begin{eqnarray} \label{eq_PPAddition}
\parity{a \boxplus b} & = & \parity{a} \oplus \parity{b} \oplus \parity{\cv{a}{b}},
\end{eqnarray}
where $c = \cv{a}{b} \in \Fq{2}^{n}$ is the so-called carry vector associated with $a$ and $b$ and obtained during the calculation of their sum. The entries of $c$ are given by
\begin{equation} \label{eq_CarryVector}
c_i = 
 \begin{cases}
  0, & \quad \text{for } i =0, \\
  a_i b_i \lor (a_i \oplus b_i) c_{i-1},&  \quad \forall i \in \inter{1}{n-1}.
 \end{cases}
\end{equation}
The addition $a \, \boxplus \, b$ of two vectors $a,b$ requires more logic gates than the XOR $a \oplus b$ and is therefore more error-prone. Hence, a variety of self-checking adders were developed such as, e.g., a parity-checked carry look-ahead adder introduced by Nicolaidis in~\cite{nicolaidis_carry_2003}.

\section{Stream Cipher ChaCha} \label{sec_ChaCha}

The ChaCha algorithm transforms a 512-bit state matrix $V \in \Fq{2^{32}}^{4 \times 4}$ 
into a unique and irreversible 512-bit output block. Encryption and decryption are performed by calculating the XOR of the keystream and the input data.  ChaCha operates on 32-bit words, and makes use of a 256-bit key $K=(\key{0} \ \key{1} \ \cdots \ \key{7})$ and a 64-bit nonce $N^\star= (\nonce{0} \ \nonce{1})$ (in several specifications 96 bits are reserved for the nonce and 32 bits for the counter).
\myalgo{
\SetKwInOut{Input}{Input}\SetKwInOut{Output}{Output}
\Input{Rounds $N \in \{8,12,20 \}$, \\
Key $K= (\key{0} \ \key{1} \ \cdots \ \key{7}) \in \Fq{2}^{128}$, \\
Counter $CTR =  (\ctr{0} \ \ctr{1}) \in \Fq{2}^{64}$, \\
Nonce $N^{\star}= (\nonce{0} \ \nonce{1}) \in \Fq{2}^{64} $.
}
\Output{Updated matrix $V \in \Fq{2^{32}}^{4 \times 4}$.}
\BlankLine
$(\const{0} \ \const{1} \ \const{2} \ \const{3}) \leftarrow (0x61707865 \ 0x3320646E \ 0x79622D32 \ 0x6B206574)$\;
$M \leftarrow$
$\begin{pmatrix} 
\const{0}  & \const{1} & \const{2} & \const{3} \\
\key{0} & \key{1} & \key{2} & \key{3} \\
\key{4} & \key{5} & \key{6} & \key{7} \\
\ctr{0} & \ctr{1} & \nonce{0} & \nonce{1}
\end{pmatrix} $; \tcp*[f]{Init.}\\
$ V\leftarrow M $\;
\For{$i\leftarrow 0$ \KwTo $N/2-1$}{
	\qr$(v_{0,0}, v_{1,0}, v_{2,0}, v_{3,0})$\; 
	\qr$(v_{0,1}, v_{1,1}, v_{2,1}, v_{3,1})$\;
	\qr$(v_{0,2}, v_{1,2}, v_{2,2}, v_{3,2})$\;
	\qr$(v_{0,3}, v_{1,3}, v_{2,3}, v_{3,3})$\;
	\qr$(v_{0,0}, v_{1,1}, v_{2,2}, v_{3,3})$\;
	\qr$(v_{0,1}, v_{1,2}, v_{2,3}, v_{3,0})$\;
	\qr$(v_{0,2}, v_{1,3}, v_{2,0}, v_{3,1})$\;
	\qr$(v_{0,3}, v_{1,0}, v_{2,1}, v_{3,2})$\;
}
$V \leftarrow V \boxplus M $; \tcp*[f]{ Entry-Wise 32-bit Sum}\\
\Return $V$\;
\caption{\texttt{ChaCha}$(N, K, CTR, N^{\star})$}\label{algo_ChaChaStateMatrix}}
 The constant $CST=(\const{0} \ \const{1} \ \const{2} \ \const{3})$, where $\const{0}=0x61707865$, $\const{1}=0x3320646E$, $\const{2}=0x79622D32$ and $\const{3}=0x6B206574$ is predefined. The counter $CTR=(\ctr{0} \ \ctr{1} )$ corresponds to the message block index $i$. In order to process the $i$-th message block, the initial state matrix $M$
is transformed in a series of $N$ rounds, where according to \cite{bernstein_chacha_2008} $N$ is suggested to be set to $8$, $12$ or $20$ (see Algorithm~\ref{algo_ChaChaStateMatrix}). The ChaCha algorithm allows to process two rows of $V$. 
In this, even-numbered rounds affect the columns of the matrix $V$, while odd-numbered rounds modify the diagonal elements of $V$.
\myalgo{\SetKwData{Left}{left}\SetKwData{This}{this}\SetKwData{Up}{up}
\SetKwFunction{Union}{Union}\SetKwFunction{FindCompress}{FindCompress}
\SetKwInOut{Input}{Input}\SetKwInOut{Output}{Output}
\Input{$a,b,c,d \in \Fq{2}^{32}$.}
\Output{$a,b,c,d \in \Fq{2}^{32}$ (Updated Values).}
\BlankLine
\BlankLine
$a \leftarrow a  \boxplus  b$; \tcp*[f]{ $\azero \leftarrow a \boxplus b$} \\
$d \leftarrow (d \oplus a)  {\lll}  16$; \tcp*[f]{ $\dzero \leftarrow d  \oplus a, \done \leftarrow \dzero  {\lll} 16 $ } \\
$c \leftarrow c  \boxplus d$; \tcp*[f]{ $\czero \leftarrow c \boxplus \done$} \\
$b \leftarrow (b \oplus c)  \lll  12$; \tcp*[f]{ $\bzero \leftarrow b \oplus \czero,\bone \leftarrow \bzero  {\lll}  12$} \\
$a \leftarrow a \boxplus b$\ \tcp*[f]{ $a' \leftarrow \azero \boxplus \bone$}\\ 
$d \leftarrow (d  \oplus a)  \lll  8$; \tcp*[f]{ $\dtwo \leftarrow \done \oplus a'$, $d' \leftarrow \dtwo \lll  8$}\\
$c \leftarrow c  \boxplus  d$\ \tcp*[f]{ $c' \leftarrow \czero \boxplus d'$}\\ 
$b \leftarrow (b  \oplus  c)  \lll \ 7$; \tcp*[f]{ $\btwo \leftarrow \bone \oplus c'$, $b' \leftarrow \btwo \lll  7$} \\
\BlankLine
\Return 
\caption{\qr$(a,b,c,d)$}\label{algo_quarterround}}
Both transformations apply the nonlinear Quarterround function shown in Algorithm~\ref{algo_quarterround}.
Each Quarterround \qr$(a,b,c,d)$ is based on four additions in $\Fq{2}^{32}$, four XORs and four rotations which operate on the 32-bit input words $a,b,c$ and $d$. A Quarterround updates each input word twice, allowing each input word to affect the other words.

\section{Parity-Based Code-Disjoint Circuit} \label{sec_Classic}

In this section we describe a parity-based code-disjoint circuit~\cite{hartje_code-disjoint_1997} for the Quarterround (Algorithm~\ref{algo_quarterround}), which is the essential part of ChaCha~\cite{bernstein_chacha_2008}. We investigate its resistance against transient and permanent faults, that can affect the input signals $a,b,c,d \in \Fq{2}^{32}$, as well as the intermediate signals $\azero,\bzero,\czero,\dzero, \bone,\btwo,\done,\dtwo \in \Fq{2}^{32}$ given in the comments of  Algorithm~\ref{algo_quarterround}.\\
For the following analysis we consider the data path of a Quarterround as illustrated on the left side of Fig.~\ref{fig_GBPPChaCha}. The right side of Fig.~\ref{fig_GBPPChaCha} shows our group-based parity prediction, which is part of Section~\ref{sec_GroupBasedPP}. The input of the data path is $a,b,c,d \in \Fq{2}^{32}$ and the output are the vectors $a',b',c',d' \in \Fq{2}^{32}$. The following four intermediate signals in $\Fq{2}$ are defined as 
\begin{equation} \label{eq_CV-QR}
\begin{split}
\alpha & \mydef \parity{\cv{a}{b}}, \quad \;\;\,  \beta \mydef \parity{\cv{c}{\done}}, \\
\gamma & \mydef \parity{\cv{\azero}{\bone}}, \quad \delta \mydef \parity{\cv{\czero}{d'}}, 
\end{split}
\end{equation}
where $\cv{a}{b}$ denotes the carry vector of $a$ and $b$ given in~\eqref{eq_CarryVector}. 
The four intermediate bits $\alpha, \beta, \gamma$ and $\delta$ defined in~\eqref{eq_CV-QR} will, in addition to $a,b,c,d \in \Fq{2}^{32}$, be used to transform a Quarterround into a code-disjoint circuit. Further on, they are used for our group-based parity approach (\texttt{GBPP}) (described in Section~\ref{sec_GroupBasedPP}).\\
A parity-based code-disjoint circuit~\cite{hartje_code-disjoint_1997} extends the classical parity prediction by additionally encoding the inputs of a given circuit into codewords of the parity code. 
Hence, we first develop a parity prediction for one Quarterround.
\begin{definition}[Parity Prediction] \label{def_PP}
Let $f$ be a function with input $x \in \Fq{2}^m$ and output $y \in \Fq{2}^n$. A parity prediction $\pp{f}$ of $f$ is a function, such that
\begin{equation*}
\pp{f}(x) = \parity{f(x)} = \parity{y}, \quad \forall x \in \Fq{2}^m.
\end{equation*}
\end{definition}
The design of the parity prediction $\pp{f}$ for a given function $f$ can be optimized in terms of, e.g., gate count and/or error coverage.

Now, we develop a parity prediction for Algorithm~\ref{algo_quarterround}, where $m = n = 128$, $x=(a \ b \ c\ d)$ and $y = (a' \ b' \ c' \ d')$ according to Def.~\ref{def_PP}. Therefore, we calculate four parity bits for each component of the output vector $(a' \ b' \ c' \ d')$.
\begin{lemma}[Parity Prediction of $a'$] \label{lem_PP-a}
Consider the output $a' \in \Fq{2}^{32}$ of a Quarterround given in Algorithm~\ref{algo_quarterround}. 
Its parity bit is 
\begin{equation*}
\parity{a'} = \parity{b} \oplus \parity{c} \oplus \parity{d}\oplus \beta \oplus \gamma.
\end{equation*}
\end{lemma}
\begin{IEEEproof}
With~\eqref{eq_PPAddition} for the output signal $a' = \azero \boxplus \bone$, we have 
\begin{eqnarray*}
\parity{a'} & = & \parity{\azero} \oplus \parity{\bone} \oplus \gamma \\
 & = & \parity{\azero}\oplus \parity{b} \oplus \parity{\czero} \oplus \gamma,
\end{eqnarray*}
and with $\parity{\czero} = \parity{c} \oplus \parity{\dzero}\oplus\beta$, we obtain
\begin{eqnarray} \label{eq_PPa-eq1}
\parity{a'} & = & \parity{\azero} \oplus  \parity{b} \oplus  \parity{c} \oplus \parity{\dzero}  \oplus \beta \oplus\gamma.
\end{eqnarray}
Inserting $\parity{\dzero} = \parity{\azero} \oplus \parity{d}$ in~\eqref{eq_PPa-eq1} leads to 
\begin{eqnarray*}
\parity{a'} & = & \parity{\azero} \oplus \parity{b} \oplus \parity{c} \oplus \parity{\azero} \oplus \parity{d} \oplus   \beta \oplus \gamma, \\
& = & \parity{b} \oplus \parity{c} \oplus \parity{d} \oplus \beta \oplus \gamma.
\end{eqnarray*}
\end{IEEEproof}

\begin{lemma}[Parity Prediction of $b', c', d'$] \label{lem_PP-allthree}
Consider the outputs $b', c', d' \in \Fq{2}^{32}$ of a Quarterround given in Algorithm~\ref{algo_quarterround}. Their parity bits are 
\begin{align*}
\parity{b'} & = \parity{a} \oplus \parity{b} \oplus \parity{c} \oplus \alpha \oplus \beta \oplus \gamma \oplus \delta, \\
\parity{c'} & =  \parity{b} \oplus \parity{d}\oplus \gamma \oplus \delta,\\
\parity{d'} & = \parity{a} \oplus \parity{c} \oplus \alpha \oplus \beta \oplus \gamma.
\end{align*} 
\end{lemma}
\begin{IEEEproof}
Similar to the proof of Lemma~\ref{lem_PP-a}.
\end{IEEEproof}

\begin{theorem}[Parity Prediction of a Quarterround] \label{theo_PP-ChaCha}
Let $(a' \ b' \ c' \ d') \in \Fq{2}^{128}$ be the output of a Quarterround given in Algorithm~\ref{algo_quarterround}. Its parity bit is 
\begin{equation*}
\pp{\qr}(a,b,c,d) = \parity{b} \oplus \parity{c} \oplus \beta.
\end{equation*}
\end{theorem}
\begin{IEEEproof}
Due to space limitations, we omit the proof.

\vspace{-.3cm}
\end{IEEEproof}
Note that the direct calculation of the parity bit of a Quarterround as given by Thm.~\ref{theo_PP-ChaCha} can be realized by 64 gates to determine the parities of the inputs, i.e., 31  XOR gates for the calculation of $\parity{b}$ (resp. $\parity{c}$), and two XOR gates to calculate $\parity{b} \oplus \parity{c} \oplus \beta$.  Another 127 gates are needed to calculate the parities of the output $(a' \ b' \ c' \ d')$ and one XOR gate to compare the parities. This results in 192 two-input XOR gates.\\
The following corollary states the circuit for transforming a Quarterround into a code-disjoint circuit as proposed in~\cite{hartje_code-disjoint_1997}. In addition, it allows to detect an odd-weight error affecting the input $(a \ b \ c \ d)$. 
\begin{corollary}[Single Output Code-Disjoint Circuit] \label{cor_CDC-ChaCha}
Let the input and output as well as the parity prediction be as in Thm.~\ref{theo_PP-ChaCha}. 
Then
\begin{equation*}
\parity{ (a \ b \ c \ d)} \oplus \pp{\qr}(a,b,c,d) = \parity{a} \oplus \parity{d} \oplus \beta.
\end{equation*}
\end{corollary}

To obtain the error coverage, we consider errors $e \in \Fq{2}^{32}\setminus \{0\}$ on the intermediate signals $\azero, \bzero, \bone, \btwo, \czero, \dzero, \done, \dtwo$ of Algorithm~\ref{algo_quarterround}. Therefore, we define the following two notations of affected vectors.
\begin{notation}[Erroneous Vector] \label{def_ErrVector}
Let $a \in \Fq{2}^{32}$ and $e \in \Fq{2}^{32}\setminus \{0\}$. 
An erroneous vector $\tilde{a}$ is defined as $\tilde{a}= a \oplus e$.
\end{notation}
\begin{notation}[Potentially Error-Affected Vector] \label{def_PotErrVector}
Let $a \in \Fq{2}^{32}$ and $e_c \in \Fq{2}^{32}$. A vector, that can be affected by an error is denoted as $\bar{c}= c \oplus e_c$.
\end{notation}
Using Notation~\ref{def_ErrVector} and \ref{def_PotErrVector} and for $c = \parity{a \boxplus b}$, the parity bit of the modulo addition as in~\eqref{eq_PPAddition} with erroneous input $\tilde{a}$ is
\begin{equation} \label{eq_PP-Err-Addition}
\bar{c} = \parity{\tilde{a} \boxplus b} = \parity{\tilde{a}} \oplus \parity{b} \oplus \parity{\cv{\tilde{a}}{b}},
\end{equation}
where ${\cv{\tilde{a}}{b}} = \cv{a}{b} \oplus e_c$. The weight of $e_c \in \Fq{2}^{32}$ can be different from the weight of $e$ (it can even be zero).

\begin{lemma}[Detectable Errors Affecting $\bzero$] \label{lem_PPError-1}
Assume an error $e \in \Fq{2}^{32}$ with odd Hamming weight is added to the intermediate signal $\bzero$ in the data path of a Quarterround (Algorithm~\ref{algo_quarterround}). 
Then, the parity prediction as in Thm.~\ref{theo_PP-ChaCha} will detect it.
\end{lemma}
\begin{IEEEproof}
The initially corrupted vector is $\tilde{\bzero} = \bzero \oplus e$ and affects $\bone$ and the output signal $b'$. Possible corrupted intermediate signals are $\gamma, \delta, \btwo,$ and the output signals $a', b',c',d'$. The parity bit of $b'$ calculated by Algorithm~\ref{algo_quarterround} is as follows:
\begin{eqnarray}
\parity{\bar{b'}} & = & \parity{\tilde{\bzero}} \oplus \parity{\bar{c'}},  \label{eq_PPError-1-1}
\end{eqnarray}
and for $\bar{d'}$, we have
\begin{eqnarray}
\parity{\bar{d'}} & = & \parity{\bar{a'}} \oplus \parity{\done}.  \label{eq_PPError-1-2}
\end{eqnarray}
Clearly, for the calculated parity we have:
\begin{equation} \label{eq_PPError-1-3}
\parity{ (\bar{a'} \ \bar{b'} \ \bar{c'} \ \bar{d'})} =  \parity{\bar{a'}} \oplus \parity{\bar{b'}} \oplus \parity{\bar{c'}} \oplus \parity{\bar{d'}}
\end{equation}
and inserting~\eqref{eq_PPError-1-1} and~\eqref{eq_PPError-1-2} into~\eqref{eq_PPError-1-3} gives:
\begin{eqnarray}
&& \parity{\bar{a'}} \oplus \parity{\bar{b'}} \oplus \parity{\bar{c'}} \oplus  \parity{\bar{d'}} \nonumber \\ 
&& \qquad = \parity{\bar{a'}} \oplus \parity{\tilde{\bzero}} \oplus \parity{\bar{c'}} \oplus \parity{\bar{c'}} \oplus \parity{\bar{a'}} \oplus \parity{\done} \nonumber \\
&& \qquad = \parity{\tilde{\bzero}} \oplus \parity{\done}. \label{eq_PPError-1-4}
\end{eqnarray}
The calculated parity \pp{\qr}$(a,b,c,d)$ as given in Thm.~\ref{theo_PP-ChaCha} is not affected by $e$. Hence, the difference between \pp{\qr}$(a,b,c,d)$ and \eqref{eq_PPError-1-4} is \parity{e} and therefore will be nonzero if $e$ has odd Hamming weight.
\end{IEEEproof}
\myalgo{
\SetKwData{Left}{left}\SetKwData{This}{this}\SetKwData{Up}{up}
\SetKwFunction{Union}{Union}\SetKwFunction{FindCompress}{FindCompress}
\SetKwInOut{Input}{Input}\SetKwInOut{Output}{Output}
\Input{$\parity{a},\parity{b},\parity{c},\parity{d}, \alpha, \beta, \gamma, \delta \in \Fq{2}$.}
\Output{$\parity{a},\parity{b},\parity{c},\parity{d} \in \Fq{2}$ (Updated Values).}
\BlankLine
\BlankLine
$\parity{a} \leftarrow \parity{a} \oplus \parity{b} \oplus \alpha$\; 
$\parity{d} \leftarrow \parity{d} \oplus  \parity{a} $\;
$\parity{c} \leftarrow \parity{c}  \oplus \parity{d} \oplus \beta$\; 
$\parity{b} \leftarrow \parity{b}  \oplus \parity{c}$\;
$\parity{a} \leftarrow \parity{a} \oplus \parity{b} \oplus \gamma$\; 
$\parity{d} \leftarrow \parity{d} \oplus \parity{a}$\;
$\parity{c} \leftarrow \parity{c}  \oplus \parity{d} \oplus \delta$\; 
$\parity{b} \leftarrow \parity{b} \oplus \parity{c}$\;
\BlankLine
\Return
\caption{\texttt{GBPP}$(\parity{a},\parity{b},\parity{c},\parity{d}, \alpha, \beta, \gamma, \delta)$}\label{algo_GBPP_RR}
}
\begin{lemma}[Detectable Errors Affecting $\czero$] \label{lem_PPError-2}
Assume an error $e \in \Fq{2}^{32}$ with odd Hamming weight is added to the intermediate signal $\czero$ in the data path of a Quarterround (Algorithm~\ref{algo_quarterround}). 
Then, the parity prediction as in Thm.~\ref{theo_PP-ChaCha} will detect it.
\end{lemma}
\begin{IEEEproof}
Similar to the proof of Lemma~\ref{lem_PPError-1}.
\end{IEEEproof}

\begin{theorem}[Error Coverage of Code-Disjoint Circuit for Algorithm~\ref{algo_quarterround}] \label{theo_ERR-PP-ChaCha}
The single output code-disjoint circuit as stated in Corollary~\ref{cor_CDC-ChaCha} for Algorithm~\ref{algo_quarterround} detects every odd-weight error $e \in \Fq{2}^{32}$ that affects
\begin{itemize}
\item[E1)] the input signals $a,b,c,d$, 
\item[E2)] the intermediate signals $\bzero, \czero,\bone, \btwo,\dtwo$, and
\item[E3)] the output signals $b', d'$.
\end{itemize}
\end{theorem}
\begin{IEEEproof}
The statement E1 follows from the properties of a code-disjoint circuit as proven in~\cite{hartje_code-disjoint_1997}. The coverage on odd-weight errors on $\bzero$ (resp. $\czero$) was proven in Lemma~\ref{lem_PPError-1} (resp. Lemma~\ref{lem_PPError-2}). From this the coverage for $\bone$ (resp. $\btwo$) follows. The coverage of odd-weight errors that affect $\dtwo$ and the output signal $d'$ is similar: The error that propagates via the addition to the output signal $c'$ is copied to $b'$ and therefore is canceled out in the sum $\parity{\bar{b'}} \oplus \parity{\bar{c'}}$, but the parity of $d'$ is affected and therefore detected. Clearly, an odd-weight error affecting $b'$ (as stated in E3) is covered, because no other output signals are affected.
\end{IEEEproof}
Note that, errors in the intermediate signals $\azero, \dzero, \done,$ and in the output signals $a',c'$ are not detected.

\vspace{-.2cm}

\section{Our Group-Based Parity Prediction} \label{sec_GroupBasedPP}

\begin{figure}[htb]
\centering
\resizebox{\columnwidth}{!}{\includegraphics{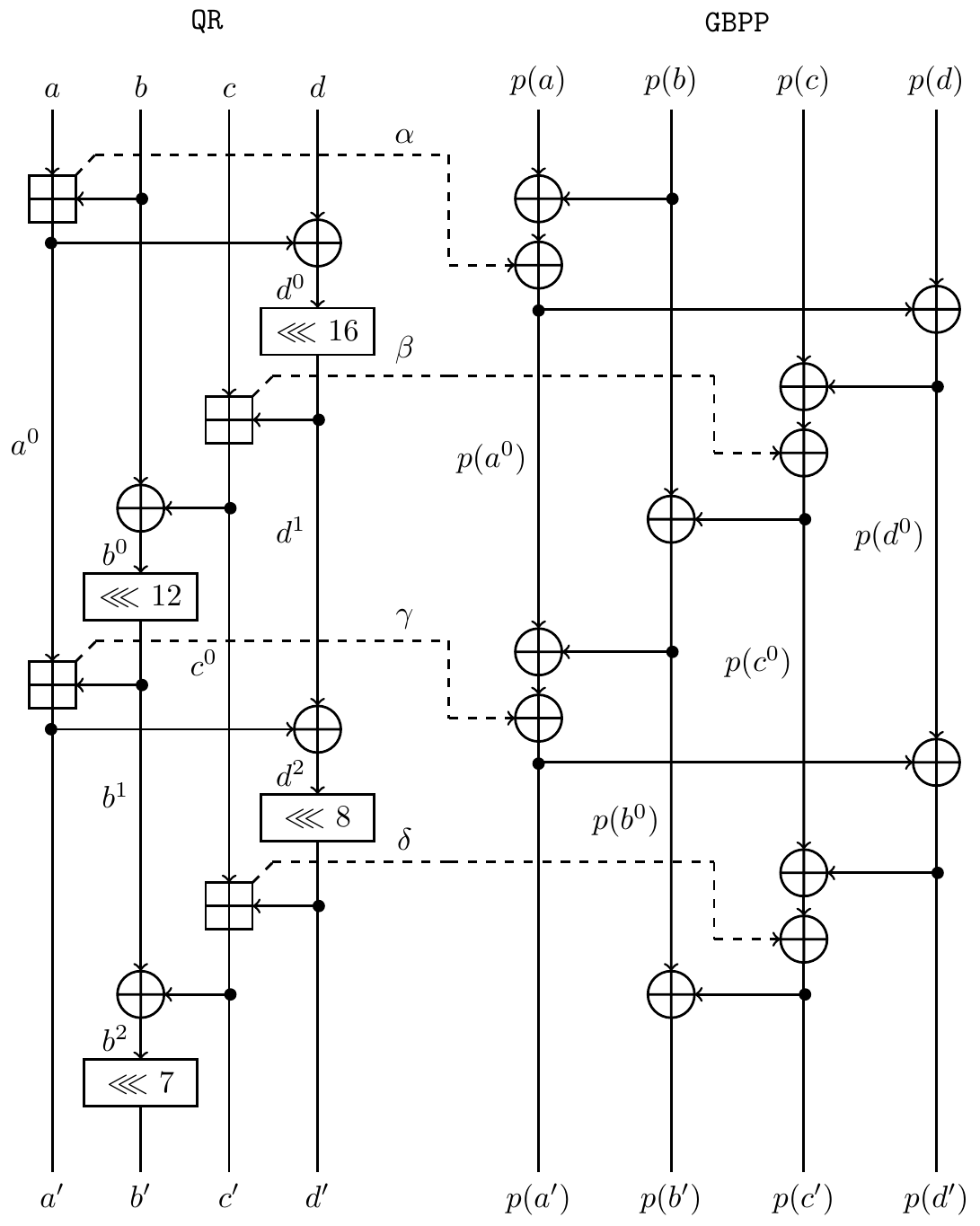}}
\caption{Our group-based parity prediction for one Quarterround of ChaCha.}
\label{fig_GBPPChaCha}
\end{figure}

To further improve the error coverage for hardware implementations of the ChaCha algorithm, we apply the method of parity prediction to the processed 32-bit words. Our approach calculates a parity bit for each of the four 32-bit components $a,b,c,d$ of the input vector of a Quarterround (Algorithm~\ref{algo_quarterround}) of ChaCha.
Fig.~\ref{fig_GBPPChaCha} illustrates our group-based parity prediction (Algorithm~\ref{algo_GBPP_RR}).
Algorithm~\ref{algo_GBPP_RR} outputs updated values of $\parity{a},\parity{b}, \parity{c}, \parity{d}$ and processes the intermediate signals $\alpha, \beta, \gamma, \delta \in \Fq{2}$ as defined in~\eqref{eq_CV-QR}
. Clearly, Algorithm~\ref{algo_GBPP_RR} is the direct translation of Algorithm~\ref{algo_quarterround} using~\eqref{eq_PPAddition} and~\eqref{eq_PPDirectSum} and the fact that the parity-bit is not changed by a bit-wise rotation (marked by ${\lll}$ in Algorithm~\ref{algo_GBPP_RR}).\\
We prove the error coverage of our approach (Algorithm~\ref{algo_GBPP_RR}) and give an estimation of the area usage in terms of required gates.

In the following, we analyze the error coverage of the proposed parity prediction scheme (Algorithm~\ref{algo_GBPP_RR}) for the intermediate signals $\azero,\bzero,\czero$ and $\dzero$ (see Lemma~\ref{lem_Error-1}, \ref{lem_Error-2}
).  The error coverage of the remaining intermediate signals, i.e., $\bone,\btwo,\done,\dtwo$ is then summarized in Thm.~\ref{theo_OddBitErrors}.
\begin{lemma}[Detectable Errors Affecting $\azero$] \label{lem_Error-1}
Assume an error $e \in \Fq{2}^{32}$ of odd Hamming weight is added to $\azero$ in the data path of a Quarterround (Algorithm~\ref{algo_quarterround}). 
Then, at least one output of our group-based parity prediction in Algorithm~\ref{algo_GBPP_RR} will detect it.
\end{lemma}
\begin{IEEEproof}
The error $e$ on $\azero$ can affect the intermediate signals $\beta, \gamma, \delta, \bzero, \bone, \czero, \dzero, \done, \dtwo, \btwo$, as well as all output vectors $a',b',c'$ and $d'$.
The initially corrupted vector is $\tilde{\azero} = \azero \oplus e$. The parity bit calculation based on the output vector $d'$ of the data path (Algorithm~\ref{algo_quarterround}) can be expressed as follows:
\begin{eqnarray}
\parityd{\bar{d'}} & = & \parity{\bar{a'}} \oplus \parity{\bar{\done}} \nonumber \\
& =& \parity{\tilde{\azero}} \oplus \parity{\bar{\bzero}} \oplus \parity{\bar{\gamma}} \oplus \parity{\bar{\done}}, \label{eq_Error-1-1}
\end{eqnarray}
and with $\bzero = b \oplus \done = b \oplus (c \boxplus \done)=b \oplus c \oplus \done \oplus \beta $, we obtain from~\eqref{eq_Error-1-1}, 
\begin{eqnarray}
\parityd{\bar{d'}} & = & \parity{\tilde{\azero}} \oplus \parity{b} \oplus \parity{c} \oplus \parity{\bar{\done}} \oplus \parity{\bar{\beta}} \nonumber \\
& & \oplus \parity{\bar{\gamma}} \oplus \parity{\bar{\done}} \nonumber \\
& = & \parity{\tilde{\azero}} \oplus \parity{b} \oplus \parity{c}  \oplus \parity{\bar{\beta}} \oplus \parity{\tilde{\gamma}}. \label{eq_Error-1-2}
\end{eqnarray}
The fourth output bit of our group-based parity prediction can be similarly expressed. In addition,  it is possibly affected by the error via $\bar{\beta}$ and $\bar{\gamma}$, i.e.: 
\begin{equation}
\parity{d'} = \parity{\azero} \oplus \parity{b} \oplus \parity{c}  \oplus \parity{\bar{\beta}} \oplus \parity{\bar{\gamma}}. \label{eq_Error-1-3}
\end{equation}
Therefore the comparison of the calculated parity $\parityd{d'}$ of $d'$ from the data path as given in~\eqref{eq_Error-1-2} and our parity prediction given in~\eqref{eq_Error-1-3} is
\begin{equation*}
\parityd{d'} \oplus \parity{d'} = \parity{e}
\end{equation*}
and will be nonzero if $e$ has odd Hamming weight.
\end{IEEEproof}

\begin{lemma}[Detectable Errors Affecting $\bzero, \czero,\dzero$] \label{lem_Error-2}
Assume an error $e \in \Fq{2}^{32}$ of odd Hamming weight is added to $\bzero$ or $\czero$ or $\dzero$ in the data path of a Quarterround (Algorithm~\ref{algo_quarterround}). 
\end{lemma}
\begin{IEEEproof}
Similar to the proof of Lemma~\ref{lem_Error-1}. 
\end{IEEEproof}

\begin{theorem}[Odd-Weight Error on All Intermediate Signals of A Quarterround] \label{theo_OddBitErrors}
Our group-based parity prediction according to Algorithm~\ref{algo_GBPP_RR} for the Quarterround (Algorithm~\ref{algo_quarterround}) detects every odd-weight error $e \in \Fq{2}^{32}$ that affects
\begin{itemize}
\item[E1)] the intermediate signals $\azero, \bzero, \czero, \dzero,\bone, \btwo, \done, \dtwo$, 
\item[E2)] the intermediate signals $\alpha, \beta, \gamma, \delta$ and 
\item[E3)] the output signals $a', b', c', d'$.
\end{itemize}
\end{theorem}
\begin{IEEEproof}
Due to space limitations, we omit the proof.
\vspace{-.2cm}
\end{IEEEproof}
The \texttt{GBPP} requires overall 265 gates. These are: 124 XOR gates for the calculation of the parity of the input words $a,b,c,d$, another $124$ XOR gates for the parity of the outputs $a',b',c',d'$, 12 additions, and 4 XOR gates in combination with a 4-input OR gate to merge the results of the four parity bit comparisons.
With the usage of fault secure adders as proposed in \cite{goessel_new_2008}, it is possible to detect any odd-weight error on input, output and intermediate signals.

\bibliographystyle{ieeetran}
\bibliography{PPnew2}

\begin{thebibliography}{10}
\providecommand{\url}[1]{#1}
\csname url@samestyle\endcsname
\providecommand{\newblock}{\relax}
\providecommand{\bibinfo}[2]{#2}
\providecommand{\BIBentrySTDinterwordspacing}{\spaceskip=0pt\relax}
\providecommand{\BIBentryALTinterwordstretchfactor}{4}
\providecommand{\BIBentryALTinterwordspacing}{\spaceskip=\fontdimen2\font plus
\BIBentryALTinterwordstretchfactor\fontdimen3\font minus
  \fontdimen4\font\relax}
\providecommand{\BIBforeignlanguage}[2]{{%
\expandafter\ifx\csname l@#1\endcsname\relax
\typeout{** WARNING: IEEEtran.bst: No hyphenation pattern has been}%
\typeout{** loaded for the language `#1'. Using the pattern for}%
\typeout{** the default language instead.}%
\else
\language=\csname l@#1\endcsname
\fi
#2}}
\providecommand{\BIBdecl}{\relax}
\BIBdecl

\bibitem{bernstein_chacha_2008}
D.~J. Bernstein, ``{ChaCha}, a variant of {Salsa}20,'' in \emph{Workshop
  {Record} of {SASC} 2008: {The} {State} of the {Art} of {Stream} {Ciphers}},
  2008.

\bibitem{bernstein_salsa20_2008}
------, ``The {Salsa}20 {Family} of {Stream} {Ciphers},'' in \emph{New {Stream}
  {Cipher} {Designs}}, ser. Lecture {Notes} in {Comput.} {Sci.}\hskip 1em plus
  0.5em minus 0.4em\relax Springer, 2008, no. 4986, pp. 84--97.

\bibitem{bernstein_poly1305-aes_2005}
------, \emph{The Poly1305-AES Message-Authentication Code}.\hskip 1em plus
  0.5em minus 0.4em\relax Springer, 2005, pp. 32--49.

\bibitem{liden_latching_1994}
P.~Liden, P.~Dahlgren, R.~Johansson, and J.~Karlsson, ``On {Latching}
  {Probability} of {Particle} {Induced} {Transients} in {Combinational}
  {Networks},'' in \emph{Proc. of {IEEE} 24th {Intern.} {Symp.} on
  {Fault}-{Tolerant} {Comput.}}, Jun. 1994, pp. 340--349.

\bibitem{goessel_new_2008}
M.~Goessel, V.~Ocheretny, E.~Sogomonyan, and D.~Marienfeld, \emph{New {Methods}
  of {Concurrent} {Checking}}, 1st~ed.\hskip 1em plus 0.5em minus 0.4em\relax
  Springer, 2008.

\bibitem{karri_parity-based_2003}
R.~Karri, G.~Kuznetsov, and M.~Goessel, ``Parity-{Based} {Concurrent} {Error}
  {Detection} in {Symmetric} {Block} {Ciphers},'' in \emph{{IEEE} {Intern.}
  {Test} {Conf.} ({TC})}, 2003, pp. 919--926.

\bibitem{bertoni_error_2003}
G.~Bertoni, L.~Breveglieri, I.~Koren, P.~Maistri, and V.~Piuri, ``Error
  {Analysis} and {Detection} {Procedures} for a {Hardware} {Implementation} of
  the {Advanced} {Encryption} {Standard},'' \emph{IEEE Trans. Comput.},
  vol.~52, no.~4, pp. 492--505, Apr. 2003.

\bibitem{natale_novel_2007}
G.~D. Natale, M.~L. Flottes, and B.~Rouzeyre, ``A {Novel} {Parity} {Bit}
  {Scheme} for {SBox} in {AES} {Circuits},'' in \emph{{IEEE} {Design} and
  {Diagnostics} of {Electr.} {Circuits} and {Syst.}}, Apr. 2007, pp. 1--5.

\bibitem{nicolaidis_carry_2003}
M.~Nicolaidis, ``Carry {Checking}/{Parity} {Prediction} {Adders} and {ALUs},''
  \emph{IEEE Trans. on Very Large Scale Integr. (VLSI) Syst.}, vol.~11, no.~1,
  pp. 121--128, Feb. 2003.

\bibitem{hartje_code-disjoint_1997}
H.~Hartje, E.~S. Sogomonyan, and M.~Gossel, ``Code-{Disjoint} {Circuits} for
  {Parity} {Codes},'' in \emph{Asian {Test} {Symposium} ({ATS} '97)}, Nov.
  1997, pp. 100--105.

\end{thebibliography}
\end{document}